\documentclass{article}
\usepackage[utf8]{inputenc}
\usepackage{graphicx}
\usepackage{hyperref}
\usepackage{authblk}

\title{Modelling remote epidemic transmission in Western Australia and implications for pandemic response}

\author[1,2,3]{Michael Small\footnote{email: michael.small@uwa.edu.au}}
\author[1]{Orlando Porras}
\author[1]{Michael Little}
\author[1]{David Cavanagh\footnote{email: david.cavanagh@integratedenergy.com.au}}
\author[1]{Harry Nicholas}

\affil [1]{Integrated Energy Pty Ltd, Como, Perth, Western Australia} 
\affil [2]{Complex Systems Group, Department of Mathematics and Statistics, University of Western Australia, Crawley, Perth, Western Australia}
\affil [3]{Mineral Resources, Commonwealth Scientific and Industrial Research Organisation, Kensington, Perth, Western Australia}

\date{June 2020}

\begin{document}

\maketitle

\begin{abstract}
We develop an agent-based model of disease transmission in remote communities in Western Australia. Despite extreme isolation, we show that the movement of people amongst a large number of small but  isolated communities has the effect of causing transmission to spread quickly. Significant movement between remote communities, and regional and urban centres  allows for infection to quickly spread to and then among these remote communities.  Our conclusions are based on two characteristic features of remote communities in Western Australia: (1) high mobility of people amongst these communities, and (2) relatively high proportion of travellers from very small communities to major population centres. In models of infection initiated in the state capital, Perth, these remote communities are collectively and uniquely vulnerable. Our model and analysis does not account for possibly heightened impact due to preexisting conditions, such additional assumptions would only make the projections of this model more dire. We advocate stringent monitoring and control of movement to prevent significant impact on the indigenous population of Western Australia.
\end{abstract}

\section{Remoteness}

The state of Western Australia (WA) comprises the western third of the Australian content. A landmass of over $2.5\times 10^6$ km$^2$, WA is not only bigger than Texas, it is larger than Texas, California, Montana, New Mexico, Arizona, Nevada and Colorado --- combined. Despite this, the population of WA is centred in the state capital of Perth. Some $92\%$ of the population of 2.6 million live in the relatively temperate South-West regions. Nonetheless, geological evidence suggests continuous habitation of WA dating back at least 120,000 years. These first peoples continue to inhabit the remote and sparsely populated landscape of much of the state. 

The current coronavirus pandemic has affected the Australian continent along with the rest of world \cite{jM20,hW20}. But, unlike much of the rest of the world, the relative remoteness of Australia and an early and robust response from government has enabled Australia to (at the time of writing) be comparatively less affected by the outbreak. Current efforts to estimate the relevant parameters for the coronavirus pandemic are currently underway globally and efforts within the Australian context are best summarised  by the technical reports of Shearer\cite{fS20}, Moss\cite{fM20}, Milne \cite{gM20} and co-workers.

Modelling specific to Australia's remote communities is somewhat less plentiful.  While chronic health problems in Australia's indigenous population are well documented and would possibly exacerbate COVID-19 related symptoms and outcome, the very small size and remoteness of many indigenous communities in Western Australia has been thought to be an asset. 

The question we set out to explore is what effect will coronavirus transmission and have on the remote communities of WA? However, this is not the exact focus of this manuscript. Here, we seek to model only spread of an hypothesised infection between urban and remote communities. Our conclusions are general and not specific only to COVID-19. To achieve this we build an agent based model of propagation on a complex network topology modelling individual contacts. We employ  expert elicitation to inform our model structure and the interaction of communities. An agent based model is developed to run on this structure with transmission parameters approximating those seen under the uncontrolled spread of coronavirus. Ensemble model predictions for a distribution of these parameter values are then used to provide an ensemble forecast of relative risk of community transmission within each community. The flexibility of the network contact model allows us to robustly test various state-level intervention strategies.

Despite a land border between WA and the rest of Australia of $1862$ km there are only two sealed roads (and one intercontinental rail line) joining WA to the rest of Australia. Connectivity between many of the regions within WA are similarly limited. During the first half of 2020 this was exploited by the state government to limit movement between the 11 administrative regions internal to the state. At the time of writing access to many remote communities within those regions (most notably, within the Kimberley) remains restricted. Nonetheless, there remains significant movement of people between these regions and settlements for a variety of traditional reasons. We model WA as an isolated entity, in reality the remote borders are somewhat porous and the conclusions of this study could be generalised to the entire Australian continent. 

The challenge of this report is how to build a useful and informative model of movement of people when data is extremely limited. For obvious reasons, mobile telephony data is limited, course-grained or unavailable, Similarly, widespread electronic monitoring and contact management has not been deployed and records of the movement of people is rather limited. 

\section{Modelling movement without movement data}

In this section we will briefly describe the data that we do have to inform this model and then describe how we use that to build an informative model. In an ideal world (from the modellers perspective) we would have information on location and movement of all $2.6$ million inhabits of WA and inferring model structure from more diverse data sources  would be unnecessary. 

\subsection{Data}

The Aboriginal and Torres Strait Islander peoples make up about $3.3\%$ of the Australian population. Collectively these people cover an incredible range of culturally diverse groups. However, identification within a particular group is often a fraught and complex process. Australian census data, collected every five years, lack the necessary level of granularity. It is estimated that there is some $120-145$ distinct languages spoken across Australia's indigenous communities. All but thirteen are considered endangered. 

For the purpose of this study we apply a grouping into $10$ distinct language clusters (of which the $10$-th is ``other''). This categorisation is based on the 2016 Census of Population and Housing and obtained from The Australia Bureau of Statistics Website. Furthermore, for each of $369$ communities ranging in size from $1.8\times 10^6$ (Perth) to $0$ (many settlements are only occupied seasonally) we record the number of speakers of each of these languages. Denote by $r_i=(r_i(1),\ldots r_i(10))$, where $r_i(j)$ is the number of speakers in community $i$ of language group $j$, a vector characterising the ``language profile'' of a given community $i$. Let $p(i)$ denote the population of community $i$. We also encode to which of $11$ regions of the state and to which local government area (LGA) each community belongs. These are absolute numbers which set the quantity of inhabitants speaking a certain language group within a community.

In Australia the LGA are often the lowest available statistical descriptor of a geographical region of statistical interest. In WA the largest LGA by area is East Pilbara and covers an area of $372,000$ km$^2$. 
 
 \begin{figure*}[t!]
\centering
\begin{tabular}{cc}
\includegraphics[width=0.325\linewidth]{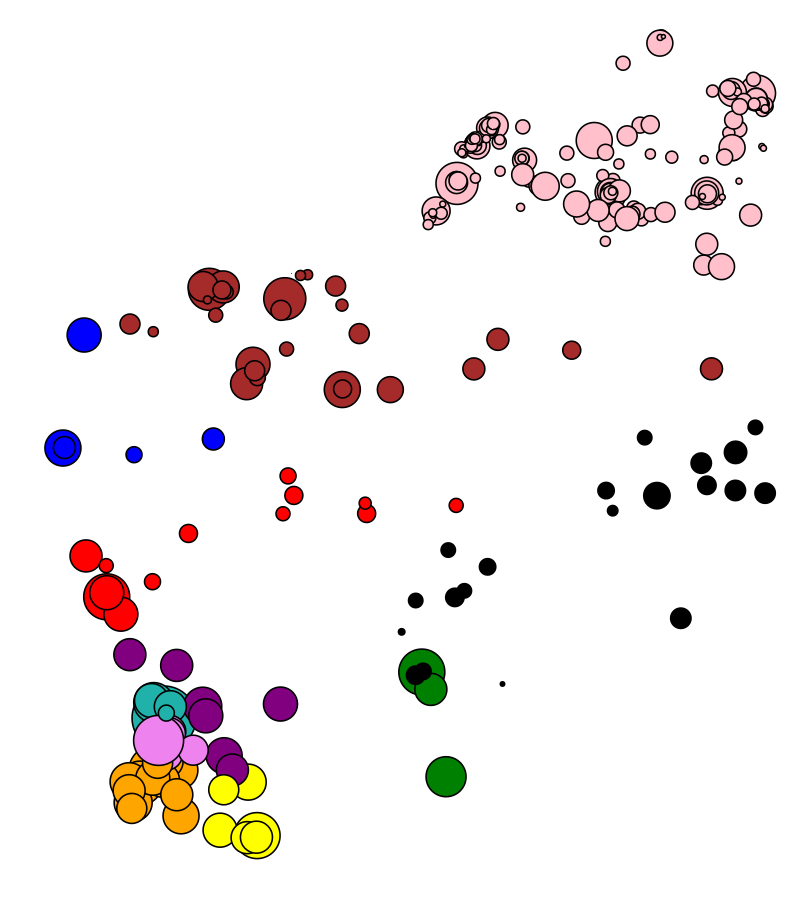}&
\includegraphics[width=0.625\linewidth]{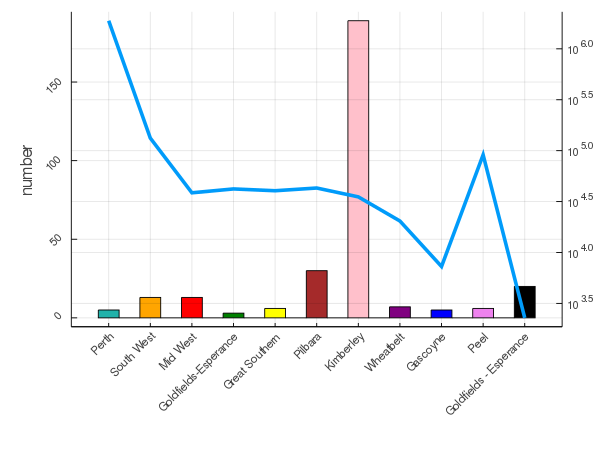}\\
(a) & (b)
\end{tabular}
\caption{Communities of WA. Panel (a) depicts the geographical location and relative size ($\log{\rm population}$) of the $297$ communities of Western Australia used in this model. Colour coding is according to region. Panel (b) indicates the number of communities in each of those regions (same colouring) and $\log{\rm population}$.}
\label{communityhist}
\end{figure*}

In the remainder of this report we consider only the $297$ communities in WA with a population $p(i)>0$. Without loss of generality we will assume that $i$ indexes only over those permanently occupied communities (i.e. $p=(p(1)\ldots p(i))$ is a vector of length 297. Figure \ref{communityhist} indicates the location within the 11 districts of WA of these $297$ communities, along with the total population of those communities.

\subsection{Model}
 
The agent based model we describe in this paper is an extension of that which we proposed in \cite{epinet}. A brief review of the most salient contributions to infection propagation models on network contact graphs is included in \cite{epinet}, and it is that approach which we follow here. Of specific note is the observation that on heterogeneous, but statistically well described contact graphs epidemic transmission may be uncontrollable \cite{rP01}. There is less work on statistical properties of transmission amongst distinct communities (in the network science sense of the word) within a graph \cite{yG18,yG19}. In summary, the approach of \cite{epinet} allows for a network based connectivity pattern to be applied across a population. Various different connection patterns are proposed in \cite{epinet} to mimic distinct contact patterns and control measures. In this report we apply the same structure at the level of communities, and then build a separate web of connections between individuals within communities.

To do this we first need to build a rule for affinity between communities --- between which communities will people travel and which communities share most in common. We model transport between communities under two regimes. First, we model the scenario (imposed by the WA government from March 31 to June 5) of no non-essential travel between regions. Subsequently, we  extend the model to allow for inter-regional travel.

\subsubsection{Regional travel ban}
\label{ban}

Travel between WA's 11 regions is only permitted if essential. To model this we assign a bonus $B_1$ if both communities are within the same region, and a second bonus $B_2$ if two remote communities are within the same LGA (urban and regional centres typically either define or span multiple LGAs and therefore $B_2$ is only used to weight connectivity between remote communities). 

Let $\delta_1(i,j)$ denote a Dirac delta-type function with value $1$ if community $i$ and $j$ are in the same region, similarly for $\delta_2$ with respect to LGAs.
\begin{eqnarray}
\label{delta1} \delta_1(i,j) &= & \left\{
\begin{array}{cc}
1 & i\ {\rm and}\ j\ {\rm are\  in\ the\ same\ region}\\
0 & {\rm otherwise}
\end{array}\right.\\
\label{delta2} \delta_2(i,j) &= & \left\{
\begin{array}{cc}
1 & i\ {\rm and}\ j\ {\rm are\  in\ the\ same\ LGA}\\
0 & {\rm otherwise}
\end{array}\right. .
\end{eqnarray}
Next we define our proxy for {\em cultural affinity} based on the dot-product of the number of speakers of each language $r_i\dot r_j$. Finally, we combine these three terms to determine the connectedness between two communities
\begin{eqnarray}
\label{similarity}
D(i,j) &=& B_1\delta_1(i,j) + B_2\delta_2(i,j) + r_i\dot r_j.
\end{eqnarray}

\subsubsection{Regional travel permitted} 
\label{noban}

To model the scenario in which regional travel is permitted we replace the affinity score (\ref{similarity}) with
\begin{eqnarray}
\label{similarity2}
D(i,j) &=& \frac{B_3}{d(i,j)} + r_i\dot r_j.
\end{eqnarray}
where $d(i,j)$ is the ordinary straight line (great circle) distance between community-$i$ and community-$j$. The parameter $B_3$ effectively weight between mass transport between communities and transport based on cultural affinity as measured through language similarity.

\subsubsection{Community affinity}

 \begin{figure*}[t!]
\centering
\begin{tabular}{cc}
\includegraphics[width=0.475\linewidth]{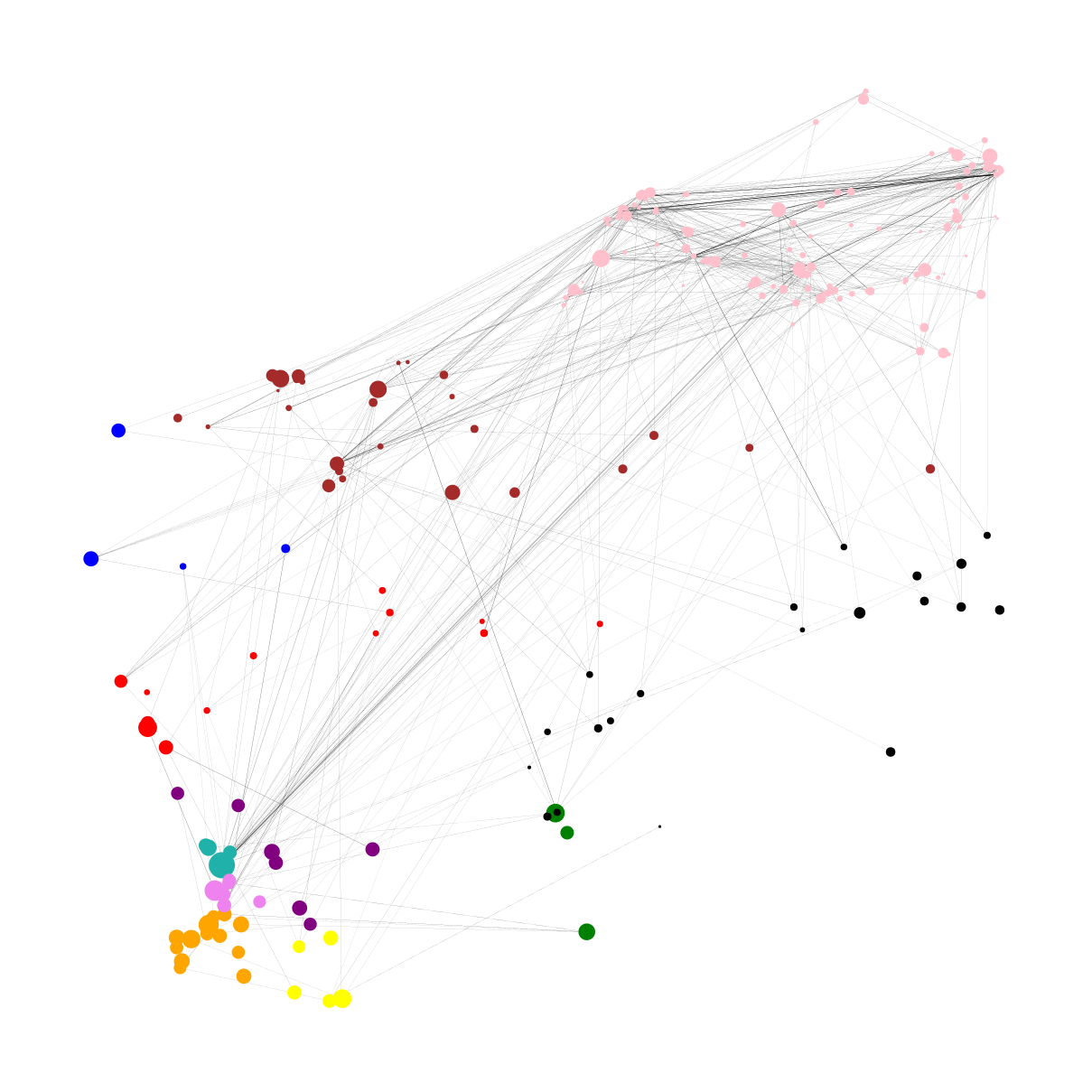} &
\includegraphics[width=0.475\linewidth]{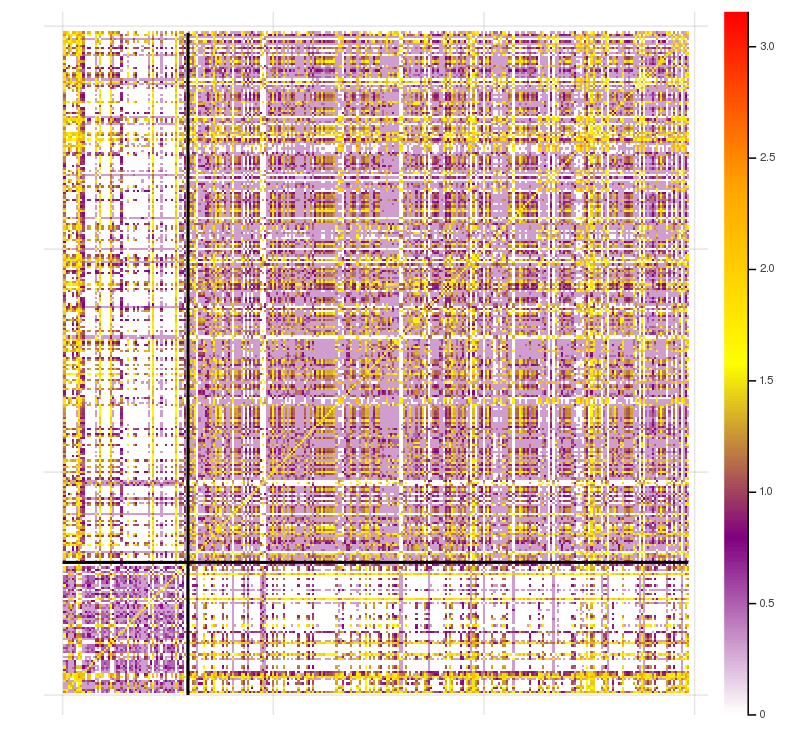}\\
(a) & (b)
\end{tabular}
\caption{Communities of WA are displayed here as a network with connectivity indicted by links (a) stronger connection indicated as darker connection and (b) the heatmap of the corresponding adjacency matrix. Communities are very approximately ordered from largest to smallest. The largest 59 towns and cities first and the large number of small remote communities subsequently. The heatmap clearly shows strong interconnection among many of the remote communities and variable connectivity strength amongst the towns and cities. The interaction between remote and urban centres is not strong giving the heatmap the block diagonal appearance.} 
\label{communitygraph}
\end{figure*}

Figure \ref{communitygraph} depicts the connectivity network which we obtain by applying this measure across the $297$ occupied communities of WA. We set $B_1=100$ and $B_2=10$. An {\em ad hoc} exploration of the parameter sensitivity of these two parameters did not show any significant consequence to this choice. Consultation with community members and Aboriginal Elders through a process of expert elicitation confirmed that the structure which we inferred was consistent with their expectations.

Note that the affinity metric $D$ is not dependent on the population of either community, but only determined by an arbitrary linear combination of two proximity measures and a measure of similarity in language. 

\subsubsection{Contact models}

We now build the connectivity matrix for the state. To do this we first decide on a connection rule from amongst those described in \cite{epinet}. In \cite{epinet} these rules were used to model the effect of different city-level control measures. We will do the same here, but the connectivity rule must be determined prior to generating the state-level contact graph. Briefly the various rules for contact within a community which were introduced in \cite{epinet} are
\begin{itemize}
    \item {\bf Scale-free network:} full mixing and no limitation of contact between individuals. Mass gatherings, crowds and movement are not restricted.
    \item {\bf Truncated scale-free network:} as above, but node degree is bounded above by $N$. Gatherings are therefore limited to no more than $N$ participants.
    \item {\bf Random graph:} random connection between nodes, but degree following a binomial distribution. Movement is permitted, but contacts are limited --- no large gatherings of any size
    \item {\bf Small-world lattice:} a small-world-like lattice structure. Movement is discouraged, non-compliance is modelled as long range connections.
    \item {\bf Lattice:} connection on a $2$-D lattice propagation is only via diffusion to neighbours. Complete lock-down, nodes only contact their (literal) neighbours.
    \item{\bf Edge thinning:} Random deletion of some proportion of links in one of the previously described network models. Models the effect of efficient contact tracing and infection identification, through, for-example, a smartphone based contact tracing application.
\end{itemize}

We choose a contact rule and apply that to each community independently and therefore build a network of $297$ disjoint components. Note that, for very small communities, each of these models is will effectively yield almost all-to-all connection. The topological structure is really only significant in large settlements. At this point we introduce a parameter $\tau$ which we characterise as the propensity to travel. In our simplest model, each community will have some proportion of travellers $\tau$ moving from that community to others. That is, $\tau P(i)$ edges emanating from community $i$ and destined to connect to other communities. 

It is likely, however, that smaller communities will be more mobile and members of smaller communities are more likely to travel --- often these very small communities will not necessarily provide all essential services. Hence let $T(i)$ denote the number of links emanating from community $i$, in the unbiased model $T_u(i)=\lfloor\tau P(i)\rfloor$. However, a biased model that allows more travel from small communities is more reasonable and under the biased model we defined
\begin{eqnarray}
\label{biased}
\label{bias}
T_b(i) &=& \lfloor \tau \sqrt{P(i)} \frac{\sum_iP(i)}{\sum_i\sqrt{P(i)}} \rfloor.
\end{eqnarray}
The choice of $\sqrt{\cdot}$ as the bias function is arbitrary, any contracting nonlinear monotonic increasing function will do.

Note that $T_i$ is the number of travellers {\em from} community $i$. Other communities may well have travellers connecting to community $i$ as well. Hence the total number of links between community $i$ and other communities will almost always be greater than $T_b(i)$ (or $T_u(i)$). In what follows we will use $T(i)$ to denote either $T_u(i)$ or $T_b(i)$.

We need to add $T(i)$ links between community $i$ and other communities. For each node we do the following. First, we choose a source node uniformly at random from amongst the nodes of community $i$. Second, we choose a destination community according to the affinity score encoded in the $i$-th row of $D$. That is,
\begin{eqnarray}
\rm{Prob}({\rm destination\ is\ community\ }j) & \propto &
D(i,j).\max\left({\frac{P(i)}{P(j)},1}\right)
\end{eqnarray}
for all $j\neq i$.
Having chosen the destination community, we need to now choose a node from within that destination community to be the target. Third, we choose that node in one of two ways: (i) uniformly at random, or (ii) with probability proportional to the degree of the target node. 

By enforcing node degree of the target node as a criteria for the destination we are selecting targets by popularity --- we are operationalising the often invoked maxim of preferential attachment in this social network. Note that we could do the same for the source node, but, only for the sake of simplicity, we choose to overlook that further complication at present.

\section{Results}

\begin{table}
\begin{center}
\begin{tabular}{c|c}\hline
$\tau$ & 0.01 \\
$B_1$ & 100\\
$B_2$ & 10\\
$B_3$ & 1000 (km)\\
\rule{0pt}{3ex}   $p$   &  $\frac{1}{5}$ \\
\rule{0pt}{3ex}  $q$   &  $\frac{1}{7}$ \\
\rule{0pt}{3ex}  $r$   &  $\frac{1}{14}$ \\
\rule{0pt}{3ex} number of seeds & $5$ \\\hline
\end{tabular}\end{center}
\caption{Epidemic simulation parameters. } 
\label{paramtable}
\end{table}

We now describe the results of the application of this model to the population of WA. In what follows, for each chosen set of parameter values and contact rule we perform $1000$ simulations with each of the $6$ distinct contact rules described above, for $210$ days each for the entire population of Western Australia. Model parameters are stated in Table \ref{paramtable}. Results reported here for the affinity rule (\ref{similarity2}) (i.e. Sec. \ref{noban}).  Results for (\ref{similarity}) naturally had significantly less transport between regions however the impact of this effect is entirely determined by the parameters $B_1$, $B_2$ and $B_3$. When computations are described for the model of Sec. \ref{ban} this will be stated explicitly.

The agent based model of disease propagation which we deploy here is the standard SEIR model presented in \cite{epinet}.
That is, at each time step:
\begin{itemize}
\item[$S\rightarrow E$] a susceptible node $i$ becomes exposed if there exists a node $j$ that is infectious (I) and $a_{ij}=1$  with probability $p$;
\item[$E\rightarrow I$] an exposed node becomes infectious with probability $q$; and,
\item[$I\rightarrow R$] an infectious node becomes removed (R) with probability $r$. 
\end{itemize}
where $a_{ij}$ is the $(i,j)$-th element of the adjacency matrix $A$ which determines epidemic transmission contacts.

 \begin{figure*}[t!]
\centering
\includegraphics[width=\linewidth]{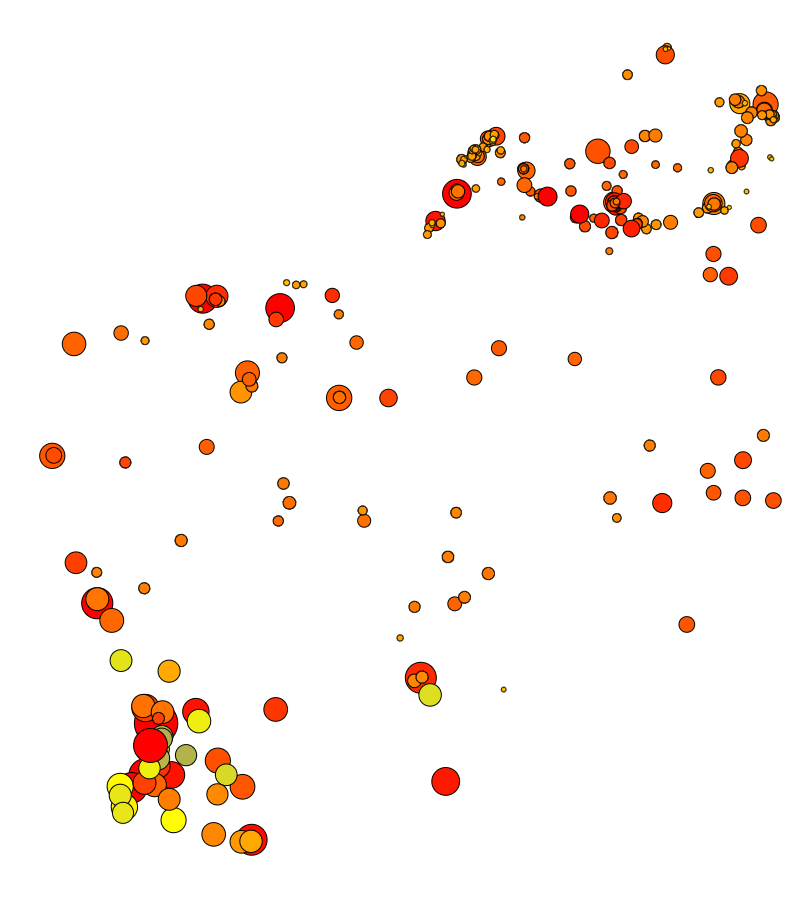}
\caption{Vulnerability of communities scaled from blue (low) to red (high). Simulations commence with exposed individuals in Perth. Note that remote communities across the Kimbereley and the outback are highly vulnerability --- more vulnerable than larger communities in the South West and closer to Perth (the largest red dot in the lower left/south west). This is due to significant interaction amongst these remote locations and also between remote and urban locations.}
\label{WA-vuln}
\end{figure*}

We quantify the vulnerability index $v(i)$ of community $i$ as follows. For each simulation run and each community we compute the number of days until local transmission is observed within that community. That gives as distribution of times until local transmission from which we extract the median $M_i$. To provide a relative measure (different simulations are under different generic conditions and need to be scaled) we compare this value to the value computed for the community of Geraldton $M_{\rm Gero}$. (a city of approximately $3.8\times 10^4$ people about $420$ km from Perth - the nearest sizeable settlement to the north of Perth). Vulnerability is then defined to be
\begin{eqnarray}
v(i) &=& \frac{M_i}{M_{\rm Gero}}.
\end{eqnarray}
Normalising against a fixed location and by computing median values we are able to robustly compare across different parameter values and amalgamate our results.

Figure \ref{WA-vuln} shows these results. The vulnerability shown here indicates very significant risk for small and remote communities in the North of the state and (relatively) low risk for larger urban areas in the South-West. Effectively, there is a large amount of transport between the remote communities and they effectively act as a single larger virtual centre. That virtual city is then strongly coupled to major cities by virtue of the fact that there is some travel to {\em each} of the remote communities. Risk across the remote communities is uniform and quite high. This is because once a community experiences local transmission, because that community is small the probability that the infected individuals then travel to other areas is extremely high. That is, the remote communities act as a strongly coupled virtual centre with the additional disadvantage that small sample sizes make the risk of further transmission very high. 

 \begin{figure*}[t!]
\centering
\includegraphics[width=\linewidth]{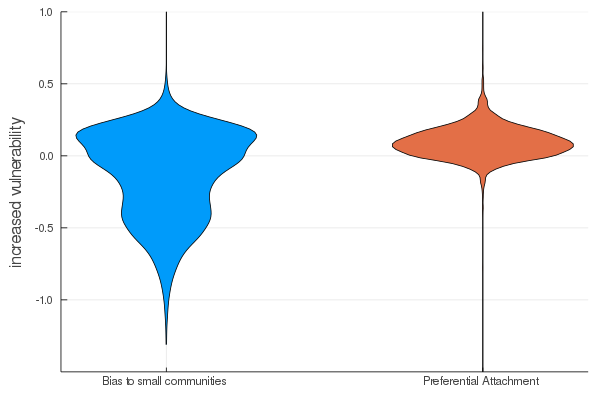}
\caption{Relative vulnerability for all communities with a bias of travel toward small communities (left) and bias to preferential attachment (right). The left violin plot shows a very slight increase in vulnerability for the majority of communities (the small communities) and a much more substantial decrease in vulnerability for the large communities when we assume more travel involves small communities. The right plot shows a slight increase in vulnerability due to preferential attachment.}
\label{WA-violin}
\end{figure*}

 \begin{figure*}[t!]
\centering
\includegraphics[width=\linewidth]{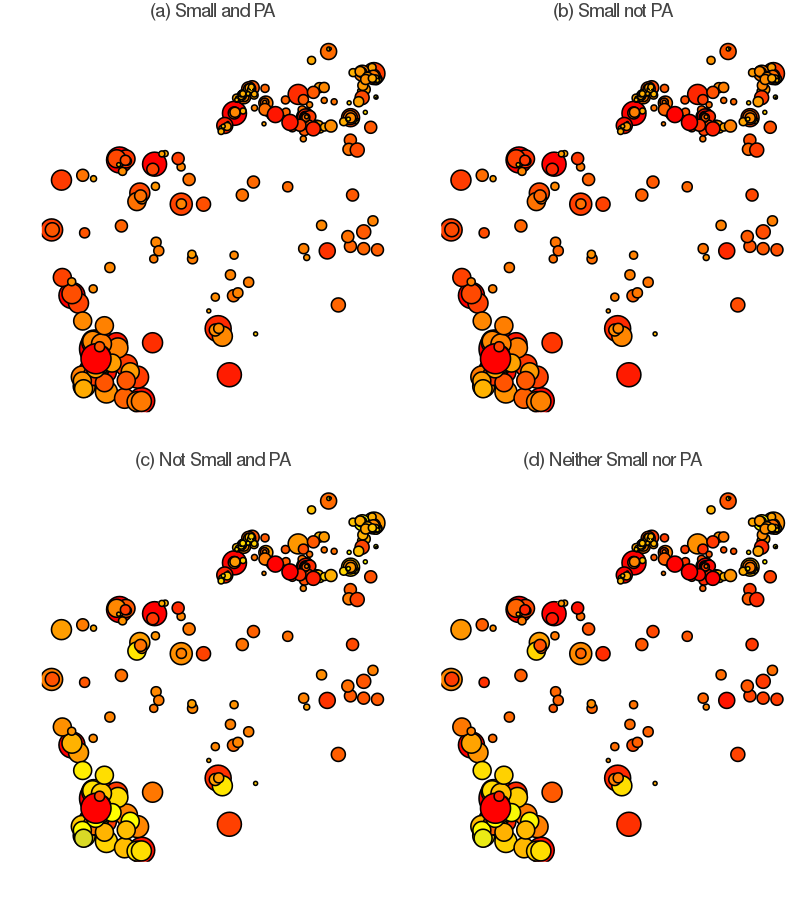}
\caption{Comparative impact of spread under the four principle modelling assumptions (a) and (b) are biased toward travel to/from small communities via (\ref{bias}) and (a) and (c) are subject to preferential attachment of travellers to higher degree nodes. Figure \ref{WA-vuln} is an average of these four plots.}
\label{4WAs}
\end{figure*}

 \begin{figure*}[t!]
\centering
\includegraphics[width=\linewidth]{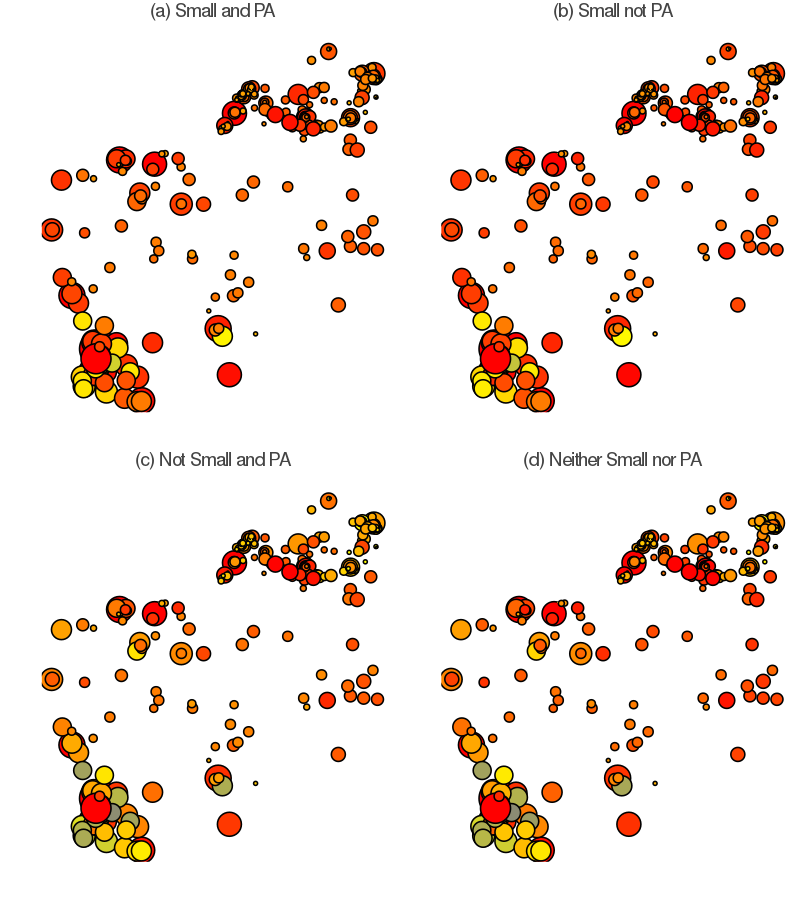}
\caption{Comparative impact of spread under the four principle modelling assumptions (a) and (b) are biased toward travel to/from small communities via (\ref{bias}) and (a) and (c) are subject to preferential attachment of travellers to higher degree nodes. This figure restricts movement according to the 11 regions of WA, movement between regions is limited (but not eliminated --- that would be trivial). Scaling reflects relative risk --- risk relative to the vulnerability of Geraldton to an outbreak in Perth. Geraldton and Perth are in separate regions.}
\label{11WAs}
\end{figure*}

 Figure \ref{WA-violin} shows the comparative effect of models assumptions related to distribution of  connections. Preferential attachment makes a slight difference --- increasing vulnerability. Unsurprisingly, the small bias term (\ref{bias}) preferencing higher mobility in smaller communities also makes a difference --- increasing vulnerability of small communities and decreasing it for large (as we would expect).  Overall though, the two effects cancel each other out (as indeed, the overall connectivity is fixed). In Fig. \ref{4WAs} the representation of Fig. \ref{WA-vuln} is repeated, but representing the four principle modelling assumptions separately. Surprisingly, preferential attachment is again seen to not have a very significant effect. We also see that bias of travel toward small communities (Fig. \ref{4WAs} panel (a) and (b) somewhat decreased the relative risk to larger communities compared to sampling travellers proportional to population (panels (c) and (d)). Fig. \ref{11WAs} repeats this computation for a simulations with internal travel restrictions between the regions of WA.

 \begin{figure*}[t!]
\centering
\begin{tabular}{c}
\includegraphics[width=0.875\linewidth]{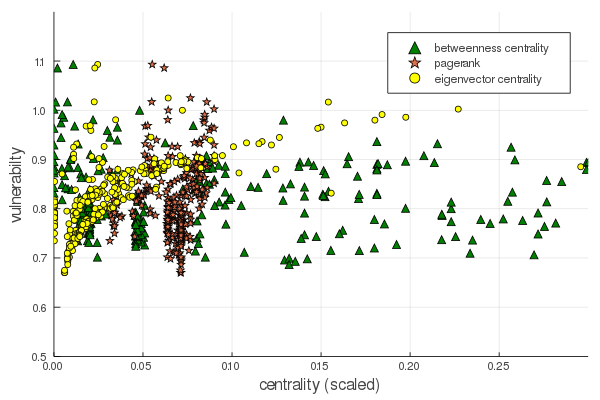} \\ (a)\\
\includegraphics[width=0.875\linewidth]{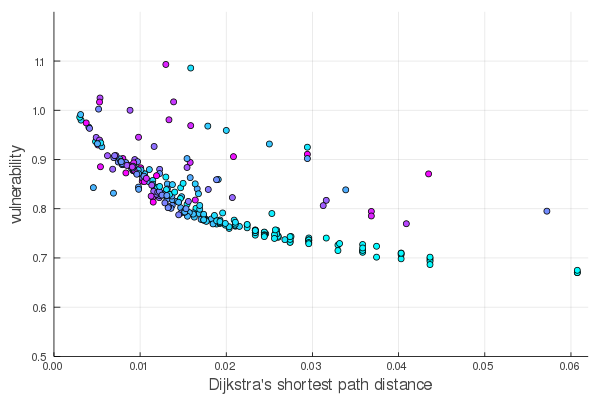}\\
(b)
\end{tabular}
\caption{Vulnerability score compared with standard network measures (a) three measures of centrality, and (b) Dijsktra's shortest path. Perth has a large vulnerability and is not shown. Node colours in (b) denote community size (larger blue and smaller purple), with the largest communities fitting the trend best.} 
\label{centrality}
\end{figure*} 

Finally, we now briefly consider the original network structure \ref{communitygraph} and whether the vulnerability deduced here could be inferred directly from the connectivity structure. Three different measures of centrality are computed and shown to have a weak dependence on vulnerability  (Fig. \ref{centrality} (a)). The most predictive measure of vulnerability is shortest path distance from Perth (Fig. \ref{centrality} (b)). Despite the fact that no microscopic connectivity information of population distribution is used, shortest path predicts vulnerability very well. Exceptions tend to be for smaller communities.

\section{Social context and Utility}

This work was developed in the context of an ongoing COVID-19 epidemic in Australia, where cases nationally are resurging particularly in Victoria.  In Western Australia a successful response included effective isolation, social distancing, hard closure of the State border, and isolation of hundreds of remote Aboriginal communities.  This was due to their heightened vulnerability from both pre-existing health conditions, and remoteness from care.  However this isolation also comes at a significant social and economic cost. 

This modelling enables the risks to remote communities to be identified, quantified, and visualized, and also provides a mechanism for the potential effectiveness of different measures to be individually assessed at a community level.  This creates a strong opportunity for communities to assess the alternative measures available to them, and the reasons behind the health guidance which is provided.  More generally, where a similar underlying dataset can be generated, this type of modelling can be extended to cases such as the potential spread of a pandemic from a capital city to surrounding regions, or surrounding states, elsewhere in Australia or the world.

\section{Conclusions}

These simulations show that remoteness and small size does not necessarily insulate communities from heightened risk of infection. Effort has been made to calibrate the parameters of this model and communities interaction to the current (2019-2020) coronavirus outbreak. Under these assumptions, and for a very wide range of other model assumptions, we have computed vulnerability of each of 297 inhabited settlements in Western Australia.  The population of these communities range over 6 orders of magnitude and yet we find that remote communities remain extremely vulnerable.

This vulnerability is borne of two key assumptions in our modelling:
\begin{itemize}
    \item A significant degree of movement between these communities, the structure of which is inferred from cultural affinity between communities. Our proxy for cultural affinity is based on prevalence of spoken language among 10 traditional language groupings.
    \item Heightened movement of resident of remote communities in comparison to urban centres. This is reasonable as services in remote communities are more limited and a smaller proportion of urban dwellers do regularly travel to remote areas. 
\end{itemize}
The remote communities are strongly coupled and therefore act as a single virtual city. Yet the distribution of contact among the population within any of these remote settlements is relatively homogeneous, and any travellers are likely to have had contact with a significant portion of other members of that community. Combined, this creates a strongly mixed virtual city that is strongly coupled to the main urban centres and the main sources of infection. Infection travels rapidly to the remote regions and very rapidly amongst them. While, in our model, it may take a significant time for community infection within a large community to pass to another town, for smaller remote communities this happens exceedingly rapidly. 

Our technical assumptions of travel bias and preferential attachment had some effect, but did not markedly change the main conclusions. Understanding of these theoretical modelling constructs was largely consistent with observed connectivity and betweenness within the similarity network. This suggest that while this simulation may be only a cartoon of a real infection outbreak, it is very robust and likely to represent true vulnerability in the communities of the Australian Outback. Moreover, network methods are able to provide a rapid assessment of expected vulnerability of communities, without the need to resort to the full simulation. Of course, this assessment is only approximate as it does not properly account for the unique structure of these small settlements.

Finally, we modelled the effect of internal border closure within the state. Of course, impermeable borders between regions would prevent infection spread. But this situation is somewhat unrealistic --- and from a modelling perspective, uninteresting. Instead, we model the situation that travel within a region is fairly easy, travel between regions is somewhat more limited --- but not eliminated. For the sake of comparison we kept overall movement fixed (the conclusion that less movement implies less transmission is hardly illuminating). Under this structure out model exhibited a patchwork structure with uniform infection across and within a particular region. Because regions of WA are hugely variable in size, this meant limited transmission in the South-West  (where there is a large number of distinct regions) but rapid transmission across large areas of the North of the state.

Our study is limited by lack of hard data of the precise internal movement of people. Nonetheless, we have tested these model conclusions across a very large number of simulations and choices of parameters. Should better data for movement become available the validity of these conclusion in a particular setting would be very easy to test.

\section*{Acknowledgements}

This work was supported by Integrated Energy Pty Ltd. MS is supported by Australian Research Council Discovery Grants DP180100718 and DP200102961. Source code for all calculations described in the manuscript is available on \url{https://github.com/m-small/epinets}.  We grateful to James Williamson (Assistant Director General, WA Health) for many helpful comments and suggestions. 

\bibliographystyle{unsrt}
\bibliography{refs}

\end{document}